\documentclass[referee]{aa}
\usepackage{epsfig}
\usepackage{lscape}
\usepackage{graphicx}
\begin{document}

\title{The field surrounding NGC 7603: cosmological or non-cosmological redshifts?}

\author{M. L\'opez-Corredoira$^1$ \and C. M. Guti\'errez$^2$}
\institute{$^1$ Astronomisches Institut der Universit\"at Basel. 
Venusstrasse 7. CH-4102 Binningen, Switzerland\\
$^2$ Instituto de Astrof\'\i sica de Canarias, E-38205 La Laguna, 
Tenerife, Spain}
\offprints{cgc@ll.iac.es}

\date{Received xxxx / Accepted xxxx}

\abstract{
We present new observations of the field surrounding the Seyfert galaxy NGC 7603,
where four galaxies with different redshifts---NGC 7603 ($z=0.029$), NGC 7603B 
($z=0.057$) and two fainter emission line galaxies 
($z=0.245$ and $z=0.394$)---are apparently connected by a narrow filament,
leading to a possible case of anomalous redshift.
The observations comprise broad and narrow band imaging and intermediate
resolution spectroscopy of some of the objects in the field. The new data
confirm the redshift of the two emission-line objects found within 
the filament connecting NGC 7603 and NGC 7603B, and settles their
type with better accuracy . Although both objects are point-like in ground based images, using
HST archive images we show that the objects have
structure with a FWHM=0.3-0.4 arcsec. The photometry in the R-band obtained during three
different campaigns spread over two years  does not show any signs of variability in
these objects above $0.3-0.4$ mag. All the above information and the relative
strength and width of the main spectral lines allow us to classify these as HII
galaxies with very vigorous star formation, while the rest of the filament 
and NGC 7603B lack star formation. 
We delineate the halo of NGC 7603 out to 26.2 mag/arcsec$^2$ in the Sloan r band 
 filter and find evidence for strong internal distortions.
New narrow emission line galaxies at z=0.246, 0.117 and 0.401 
are also found at respectively 0.8, 1.5 and 1.7 arcmin to the West of the 
filament within the fainter contour of this halo.
We have studied the spatial distribution of objects in the field within 
1.5 arcmin of NGC 7603. 
We conclude that the density of QSOs is roughly within the 
expected value of the limiting magnitude 
of our observations. However, the configuration of the four 
galaxies apparently connected by the filament 
appears highly unusual.
The probability of three background galaxies of any type with 
apparent B-magnitudes up to 16.6, 21.1 and 22.1
(the observed magnitudes, extinction correction included) being randomly 
projected on the filament of the fourth galaxy (NGC 7603) 
is computed resulting $\approx 3\times 10^{-9}$. 
Furthermore, the possible detection of 
very vigorous star formation observed in the HII galaxies of the filament 
would have a low probability if they were background normal-giant galaxies; 
instead, the intensity of the lines is typical of dwarf HII galaxies.
Hence, a set of coincidences with a very low probability
would be necessary to explain this as a fortuitous projection of background sources. 
Several explanations in terms of cosmological or non-cosmological redshifts are discussed.
\keywords{Galaxies: individual: NGC 7603 --- 
Galaxies: statistics --- Galaxies: peculiar --- Galaxies: starburst --- 
distance scale}}

\authorrunning{L\'opez-Corredoira \& Guti\'errez}

\titlerunning{NGC 7603}

\maketitle

\section{Introduction}

\subsection{Anomalous redshift problem}

The problem of the apparent optical associations of galaxies with very different redshifts, the
so-called anomalous redshifts (Narlikar 1989; Arp 1987, 1998), is old but still alive. 
Although surprisingly ignored by most of the astronomical 
community, there is increasing evidence of examples of such anomalies. 
Statistical evidence has grown for such associations over the last 30
years (Burbidge 1996, 2001). For instance, all non-elliptical galaxies brighter than 12.8 mag
with apparent companion galaxies have been examined (Arp 1981), 
and 13 of the 34 candidate companion galaxies were found to have QSOs with higher redshift.
Given an accidental probability of less than 0.01 per galaxy,
the global probability of this to be a chance is $\sim 10^{-17}$.
Bias effects alone cannot  be responsible for these correlations
(Burbidge 2001; Hoyle \& Burbidge 1996; Ben\'\i tez et al.  2001).
Weak gravitational lensing by dark matter has been proposed
(Gott \& Gunn 1974; Schneider 1989; Wu 1996; Burbidge et al. 1997) as the cause of these 
correlations, although this seems to be insufficient to explain them
(Burbidge et al. 1997; Burbidge 2001; Ben\'\i tez et al. 2001;
Gazta\~naga 2003; Jain et al. 2003),
and cannot work at all for the correlations with the brightest and
nearest galaxies. The statistical relevance of these associations
is still currently a matter of debate 
(Sluse et al. 2003).

A recent compilation of associations galaxies$-$QSOs has been presented by 
Burbidge (1996). Some remarkable cases of apparent associations between 
objects with different redshift are Arp 220 (Ohyama et al. 1999; Arp et al. 2001), 
NGC 1068 (Burbidge 1999a; Bell 2002a), NGC 3067 (Carilli et al. 1989;
Carilli \& van Gorkom 1992), NGC 3628 (Arp et al. 2002), NGC 4258 (Pietsch et al. 1994;
Kondratko et al. 2001), NGC 4319 (Sulentic \& Arp 1987), etc. Some of these may be just fortuitous 
cases in which background objects are close to the foreground galaxy, 
although the statistical mean correlations remain to be explained, and some cases
alone have very small probability to be a projection of background objects. 

Associations of galaxies with different redshifts might also take place: 
forty-three systems among the hundred Hickson (1982) groups of galaxies 
(compact groups of galaxies containing four to six members) have one redshift very 
different from the mean of the others (Sulentic 1997). 
For instance, Stephan's quintet (Moles et al.
1998; Guti\'errez et al. 2002), the chain VV172 (Arp 1987; Narlikar 1989), etc.
Although the numbers, sizes, magnitudes and morphological
types of the discordant redshift members might agree with a scenario
of chance projections, the distribution of
positions in quintets is more centrally concentrated than expected in
such a scenario (Mendes de Oliveira 1995). This author claims
that compact groups might act as gravitational lenses and therefore explain
the difference in concentration, but this remains to be justified. 

To explain these associations Hoyle et al. (1993)
 proposed new physics in which part of the measured redshifs are not caused by 
the expansion of the Universe. Other theories have been proposed too (see \S \ref{.discussion}).
We are carrying out a series of observations
of some of the suspicious systems in order to know more about them and
to throw further light to the problem (Guti\'errez et al. 2002; 
L\'opez-Corredoira \& Guti\'errez 2002; Guti\'errez \& L\'opez-Corredoira 2003 in
preparation;
and this paper). In particular, this paper is about the system of NGC 7603 and 
the surrounding objects.

\subsection{NGC 7603}

The main galaxy, NGC~7603, is a broad line Seyfert I galaxy 
with z$=0.0295$ and $B=14.04$ mag (de Vaucouleurs et al. 1991).
This galaxy has been studied mainly  in relation to its variability, which 
 was discovered by Kopylov et al. (1974), and Tohline \& Osterbrock
(1976). Kollatschny et al. (2000) have presented the results of an extensive
programme to study  the line and continuum variability over a period of twenty
years. They detected spectral variations on timescales from months to
years. The variability observed is $5-10$ in the intensity
in the Balmer and Helium lines, and in
the continuum. From the perspective of the Eigenvector 1 parameter space 
for AGNs (Sulentic et al. 2000, 2002), the Balmer lines are unusually broad 
and show a very 
complex structure. The Balmer lines are blueshifted relative to the 
local `rest frame' of the AGN by between 1000 and 2000 km/s. Less 
than 5\% of AGN show such characteristics. Such lines are more common in 
radio-loud quasars, where one sees ejected synchrotron lobes. It shows 
unusually strong FeII emission for an AGN with such broad 
lines (Goodrich 1989, Kollatschny et al. 2000).

The system around NGC 7603 is very interesting because it is among 
the cases (Arp 1980) with some filamentary structure 
joining galaxies with different redshift. Arp (1971, 1975, 1980) 
has claimed that the compact member has somehow been ejected from the bigger object.
NGC 7603 and its filament are so distorted that significant tidal 
disturbance can be reasonably assumed, without a clear candidate for
companion galaxy producing the tides (see \S \ref{.diffuse}).
Another fact that has attracted attention (Arp 1971, 1975; Sharp 1986)
is the proximity of NGC~7603B, a spiral galaxy with higher redshift (z$=0.0569$)
located 59 arcseconds to the SE of NGC 7603. 
The angular proximity of both galaxies and the apparently luminous
connection between them has converted the system into an important example of
an anomalous redshift association. Hoyle (1972) has pointed out that
NGC 7603 is one of the most strange cases, and which needs a non-standard theory
to be explained. Apart from the above facts there
are also two in principle point-like objects superimposed on the filament
that apparently connects both galaxies. We thought that the particularities of
the systems deserve more attention and decided to study the bridge and the
point-like objects mentioned. 

In L\'opez-Corredoira \& Guti\'errez (2002,
hereafter Paper I) we presented intermediate resolution spectra of the
filament and the two objects mentioned  (see Fig. 1 of Paper I).
From several absorption lines we estimated the redshift of the filament 
apparently connecting NGC 7603 and NGC 7603B as z$=0.030$, very
similar to the redshift of NGC 7603 and probably associated with this galaxy. 
We identified several emission lines
in  the spectra of the two knots and from the emission lines of H$\beta$, OII
(3727\AA) and OIII (4959 and 5007 \AA) we determined their
redshifts, obtaining 0.39 and 0.24 for the objects closer and farther from NGC
7603 respectively. The two objects might be QSOs or HII-galaxies.
The spectra of paper I had not enough resolution to determine their nature definitively 
 (since we used a wide slit) and the seeing conditions limited the
possibility of seeing structure under 1 arcsecond in these objects.
These intriguing results motivated us to continue with the
study of this system. We planned new observations with the aim of
answering the following questions: i) What is the nature of the two knots in the
filament?, ii) Are there any other high redshift objects in the halo 
surrounding NGC 7603?, and iii) Are there any clues in the surrounding field that 
 help us understand the nature of this apparent association?

This paper contains the analysis of new observations and is structured
as follows: Section \ref{.observ} presents the details of the observations and 
data reduction. Section \ref{.imaging} presents the observed
images, and the main features discovered in each component. 
Section \ref{.spectra} presents the spectroscopy of some sources.
Section \ref{.analysis} calculates the probabilities of the observed 
configuration to be an accidental projection of background galaxies,
and discusses the results presenting some possible
physical scenarios to explain them. A summary with the main results is given 
in Section \ref{.conclus}.

\section{Observations}
\label{.observ}

The observations presented here comprise narrow and broad band imaging, and
spectroscopy with intermediate resolution. These observations were taken at
the IAC80\footnote{The telescope is at the Spanish Teide Observatory on the
island of Tenerife and is operated by the Instituto de Astrof\'\i sica de
Canarias}, NOT\footnote{The Nordic Optical Telescope (NOT) is operated on the
island of La Palma jointly by Denmark, Finland, Iceland, Norway, and Sweden, in
the Spanish Observatorio del Roque de los Muchachos of the Instituto de
Astrof\'\i sica de Canarias.}, WHT\footnote{The William Herschel Telescope
(WHT) is operated on the island of La Palma by the Isaac Newton Group of Telescopes.}
telescopes, and from the HST archive\footnote{Based on observations made with the NASA/ESA 
Hubble Space Telescope, obtained from the data archive at the Space Telescope Science 
Institute. STScI is operated by the Association of Universities for Research in Astronomy, Inc. 
under NASA contract NAS 5-26555.}. 
Table \ref{sum_table} presents a summary of the observations.  


\begin{table*}
\caption{Observations.}
\label{sum_table}
\begin{center}
\begin{tabular}{lllll}
Telescope and instrument & Epoch & Mode & Exposure time & Seeing \\
\hline
IAC80 (0.8 m) CCD  & Jul.8-00/Aug.17-01 & Image, narrow-fi.IAC39-6767 \AA & 30000 s (dark) & 1.8" \\
		 & Aug.6-00/Jul.22,Aug.15-01 & Image, narrow-fi.IAC35-6931 \AA & 52834 s (grey)  & 1.8" \\
NOT (2.6 m) ALFOSC & 2000 June 13 & Image, narrow-fi.IAC39-6767 \AA  & 900 s (dark)		   & 1"  \\
		 & 2000 June 13 & Image, narrow-fi.IAC35-6931 \AA  & 900 s (bright)	   & 1" \\
 	         & 2000 June 13 & Image, R-Bessel band   & 900 s (dark)	   & 1" \\
     	         & 3001 August 12 & Image, R-Bessel band   & 300 s (dark)		   & 1.5"-2.0" \\
     	         & 2002 November 30 & Image, R-Bessel band   & 900 s (dark)		   & 1.5"-2.0" \\
     	         & 2002 November 30 & Image, u-Sloan band   & 3200 s (dark)		   & 1.5"-2.0" \\		 
     	         & 2002 November 30 & Image, g-Sloan band   & 1200 s (dark)		   & 1.5"-2.0" \\		 
     	         & 2002 November 30 & Image, i-Sloan band   & 900 s (dark)		   & 1.5"-2.0" \\
     	         & 2002 December 2 & Image, r-Sloan band   & 1200 s (non-phot.)   & 1.5"-2.0" \\
     	         & 2002 December 3 & Image, r-Sloan band   & 4200 s (dark)   & 1.5"-2.0" \\
\ \ \ \ \ \ \ (only Paper I) $-->$  & August 12, 2001 & Spectr.l.slit, gr.\#4,
ap.5"  & 14225 s (dark)	   & 1.5"-2.0" \\		 
WHT (4.2 m) ISIS red arm & 2002 December 28  & Spectr.l.slit, gr.R158R, ap.1.2"  & 5400 s (Pos 1) & --- \\
	               & 2002 December 29 & Spectr.l.slit, gr.R158R, ap.1.5"  & 5400 s (Pos 2) & --- \\
                       & 2002 December 29 2002 & Spectr.l.slit, gr.R158R, ap.1.5"  & 1800 s (Pos 3) & --- \\
 HST WFPC2 & 1994 July 3 & Image, filter F606W & 500 s.	   & ---  \\
\end{tabular}
\end{center}

\end{table*}


We wanted to check for the presence of H$\alpha$ emission in the filament
connecting NGC 7603 and NGC 7603B as well as in the galaxies themselves.  NII
(6583 \AA) is also interesting and might be stronger  if e.g. shocks were
involved; it would be observed  in narrow filters of FWHM 50 \AA \ centered at
H$\alpha $ emission. During several campaigns in 2000 and 2001 we obtained
imaging at the IAC80 and NOT  with the  IAC39 and IAC35 filters which are
centred on 6767 and 6931 \AA~and which match the H$\alpha$ line at velocities
of 9,372 and 16,870 km$\, $s$^{-1}$ respectively and have a FWHM equivalent to
2,000 km$\,$s$^{-1}$. These ranges in velocity correspond to the redshifts of
NGC 7603 and NGC 7603B respectively. The images were reduced using a standard
procedure that comprises bias subtraction, flat-field correction, shifting and
co-addition of individual exposures. The continuum in each case was subtracted
using a resampled and scaled image  (in order to have the same resolution of
the IAC80: 0.435 arcsec pixel$^{-1}$) in the R band taken on 2000 June 13th at
the NOT.

With the broad band images we wanted to delineate in detail the halo of the
system NGC 7603 - NGC 7603B, to detect other possible candidates in the field
and measure their colours,  and to constrain possible variability of the two
objects in the filament.  For the study of the variability we took several
images in the R band (Bessel) in the period 2000-2002.  For the remaining
tasks, apart from the R filter, we observed the field with the Sloan u, g, r,
and i  filters.  In all cases we used the NOT  with the ALFOSC instrument.
The images were reduced following the standard procedure mentioned above. The
conditions were photometric in all runs except on December 2.  For the 2000
June 13th  observation, several Landolt calibration fields (Landolt 1992)
were observed. The observations in the other two runs with the R filter at
the NOT (2001 August 12th and 2002 November 30) were relative calibrated with
respect to this using  using eight stars in the field. For the Sloan filters,
we have calibrated with some stars from the list given by Smith et al.
(2002).

Because of limiting atmospheric conditions, we could not see details
below 1 arcsec  resolution from the ground telescopes images. Therefore,
Hubble Space Telescope archive were also used to obtain a high spatial resolution of the objects
embedded in the filament. These data come from the HST Proposal 5479 made by 
Matthew Malkan, which was used to produce the paper Malkan et al. (1998).
The image, although less deep (exposure time: 500 s.), 
allows us to see small scale details of some interesting objects, 
since this includes the filament connecting NGC 7603 and NGC 7603B.

We obtained spectroscopy in two campaigns, the first at the NOT  using
ALFOSC (presented in Paper I and not considered here)
and the second  nearly a year and a half later 
using ISIS at the WHT in order to get further and better spectra than in Paper I
and to study other objects in the field. At the WHT we put the slit in three different
positions to optimize the observation of the objects within the filament and 
several other objects that were selected according to their colours (see \S
\ref{.colours}). 
The grism used was R158R. We took Tungsten, and Cu-Ne and Cu-Ar
calibration lamps for flat-field correction and spectral calibration
respectively. The data were bias subtracted. After some tests we decided not
apply any flat fielding correction because such corrections would require prohibitive exposure
times with the Tungsten lamp on the blue side of the spectrum and in any case 
this correction is very small ($\sim 1$\%) in the red part of the spectrum. 
The FWHM measured of the lines is 8 \AA \ for the first position,
and 20 \AA \ for the second and third position.
We extracted the spectra using the task
$apall$ of IRAF\footnote{IRAF is Image Reduction and Analysis Facility,
written and supported by the IRAF programming group at the National Optical
Astronomy Observatories (NOAO) in Tucson, Arizona.}. The data are sampled at 
1.62 \AA/pixel and 
covers from 2810 \AA~to 10450 \AA. However, due to the response of the grism, 
the sensitivity of the first 1000 pixels (below 4400 \AA) or the last 700 pixels
(over 9300 \AA) is very poor and have not been used in any of the analysis.

\section{{\bf Imaging}}
\label{.imaging}

\subsection{{\bf Morphology and surface photometry}}
\label{.diffuse}

Figure~1 shows the R band image obtained combining the different observations
in this band (see also Fig. 1 of Kollatschny et al. 2000).  The figure presents the
grey-scale and isophotal maps in this filter. The high emission due to the activity of
the galaxy NGC 7603 saturates the image in the central part of this galaxy. The system
NGC 7603--NGC 7603B appears to be surrounded by a diffuse halo that we have been able to
delineate out  to 26.2 mag/arcsec$^2$ in the Sloan r-band  filter. Although this halo 
seems to be associated mostly with NGC 7603,  it is not symmetric with respect to this
galaxy. There is evidences of a fainter extension tail in northern direction. The last
isophote of the halo is also asymmetric  to the West, possibly including a counter arm of
the bright filament between NGC 7603 and NGC 7603B.  The halo+filament between NGC 7603
and NGC 7603B shows up clearly and has a maximum brightness of 22.9 mag/arcsec$^2$ in the Sloan
r-band   filter, while the halo has a brightness near the filament of 23.4 mag/arcsec$^2$.
Therefore, the filament alone has around 24.0 mag/arcsec$^2$. Another diffuse structure
is seen apparently connecting  NGC 7603  and NGC 7603B also, and situated to the South of
the main filament. A point like object (\#17 of Fig.~4)  situated in
the southest point of this tail has been also observed   spectroscopically (see below)
resulting a local star. 

Fig.~2 shows a map of  contours of the H$\alpha$ emission in the 
IAC39 filter (centred on the redshift of the galaxy NGC 7603) once the continuum
(R-filter) is roughly subtracted. No emission was found in the IAC35 filter
(centred on the redshift of the  galaxy NGC 7603B), either in  NGC 7603B or in
the filament. Only the nucleus of the NGC 7603 (IAC39 filter) shows some
emission, as expected from a  Seyfert 1 galaxy. No stripped emission regions (as
found, for instance, in the stripping event in Stephan's quintet; Sulentic et
al. 2001, Guti\'errez et al. 2002) were observed. This absence of H$\alpha $
emission lines in NGC 7603B has already been pointed out by Sharp (1986). The
non-detection of emission lines is not proof against the existence of a physical
connection. In interactions and ejections with a larger galaxy, the gas is often
stripped out of a stellar system (Rose et al. 2001); so the lack of emission
lines could be taken as an indication of interaction rather than non interaction
(pointed out by Sharp 1986).

Figs.~1, 2 and 3 show that
NGC 7603 and its filament are apparently distorted by significant tidal interaction.
The own existence of the filament is also a possible sign of tidal interaction or a debris from satellite
disruption (Johnston et al. 2001). The fainter southern filament (the one which crosses 
object \#17) and the red fringe embedded in NGC 7603 (red colour in
Fig.~3; due possibly to dust) reinforces the scenario with tidal debris. The colour of
the filament connecting NGC 7603 and NGC 7603B is an average one: $(g-r)\approx 0.95$ (equivalent 
to $(B-V)\approx 1.15$, like, for instance, 
in the bridge of the interacting system Arp 96; Schombert et al. 1990), $(u-g)\approx 1.5$,
like the outer region of NGC 7603, and there are not emission lines in the filament (Paper I).
The colour of this filament is relatively blue compared to
the average value in the sample of tidal features by Schombert et al. (1990).

\subsection{{\bf The Neighborhood of NGC 7603}}
\label{.colours}

First, we looked for QSOs, since they are typical objects among anomalous redshift
candidates.
We try to identify QSOs with $z<2.5$ in the field using the multicolour 
criteria proposed in the analysis of the 2dF Survey (Boyle et al. 2000; Meyer et al. 2001). 
This criterion, converted into Sloan filters through the relations between the 
UBVRI Johnson  filters
and ugri Sloan filters given by Smith et al. (2002), and the relation between the photographic filter
$b_j$ and Johnson filters: b$_j$=B-0.28(B-V) (Meyer et al. 2001); (we adopt the approximation
of U, R photographic-filters equivalent to U, R Johnson-filters) is:

\begin{equation}
(g-r)<
\left \{ \begin{array}{ll} 
        \frac{0.552-(u-g)}{0.381},& \mbox{ $(u-g)<0.452$ } \\ 
        \frac{0.921-(u-g)}{1.793},& \mbox{ $0.452<(u-g)<1.197$ } \\ 
        -0.154,& \mbox{ $(u-g)>1.197$ } \\ 
\end{array} 
\right \} 
\label{selecQSO}
.\end{equation}

The completeness is quite high;  this criterion covers $80-90$\% of all QSOs (Meyer et
al. 2001). The principal contamination comes  from Galactic stars, subdwarfs and white
dwarfs, and blue compact emission-line galaxies (Croom et al. 2001). The total fraction
of the contaminant sources is $\approx 45$\% (Croom et al. 2001). Therefore, the total
number of QSOs will be $\approx 2/3$ times the number of  objects that follow the
criterion of eq. (\ref{selecQSO}).  This number is slightly different when  the range of
magnitudes is different from those of the 2dF survey, but not by too much.

We proceed as follows: first we select all objects in the field detected in the u filter. 
There are 38 objects in the u filter, including the two knots in the filament, but excluding 
NGC 7603 and NGC 7603B.
We used the software ``Sextractor'' (Bertin \& Arnouts 1996) to measure the photometry
of these objects in u, g, r, i filters. For the two objects in the filament first we tried to subtract the
contribution of the filament by a two-dimensional surface fit. Although we 
tried with different functions, ranges, etc., the result was not perfect
satisfactory partly owing  to the presence of the two main galaxies, but the accuracy in the estimation of the magnitudes
for these two objects is good enough within an uncertainty of $\sim 0.2$ mag.
Also, the photometry of object \#35 was done separately because it was
embedded in the halo of NGC 7603, taking special care of sky subtraction. It is 
noteworthy that  object \#35 has a quite high value of (g-r)=1.8,
while (u-g)=0; although it does not follow eq. (\ref{selecQSO}), it
may be a unusual object because of its  colours.
Table \ref{ugri_table} presents the results on the
photometry in u, g, r and i of all the objects, whose positions are shown in
Figure~4. 
We have no u magnitude for object \#1 because it is too faint in this filter.
Figure~5 presents a colour-colour diagram for these objects, and the regions in 
which QSOs are expected. We see that objects \#19, \#23 and \#36 follow the criterion of eq.
(\ref{selecQSO}), which is indicated in Figure~5.

{\bf
\begin{table}
\caption{Magnitudes of the objects in the field of NGC 7603 derived
using ``Sextractor'' (except those marked with *, which were derived
separately with ``phot'' taking care of the filament/halo subtraction
and are affected by an error of at least 0.2 mag.). Last column points out
whether they are extended ``E'' (as far as we can see from the available images
and spectra; some furhter faint objects which look point-like might be extended too).}
\label{ugri_table}
\begin{center}
\begin{tabular}{rrrrrr}
\# &	u	&   g	   &   r 	&   i	&  Ext. \\
\hline
1  &   ----	&   23.1(*) &     21.7(*) &   21.6(*) & E \\
2  &   23.8(*)	&   22.9(*) &     22.2(*) &   21.8(*) & E \\
3  &   20.7	&   19.4 &     18.7	&   18.6 & E	\\
4  &   20.9	&   18.1 &     16.6	&   16.0	\\
5  &   21.8	&   19.7 &     18.9	&   18.8	\\
6  &   18.6	&   17.3 &     16.7	&   16.7	\\
7  &   22.6	&   21.0 &     20.4	&   20.0 & E	\\
8  &   17.6	&   16.0 &     15.3	&   15.1	\\
9  &   22.9	&   21.1 &     20.5	&   20.2 & E	\\
10 &    21.5	&   21.2 &     20.4	&   20.5	\\
11 &    23.7	&   21.8 &     20.5	&   20.2	\\
12 &	23.3	&   22.4 &     21.3	&   21.2	\\
13 & 	 19.7	&   18.6 & 	 18.1	&     18.2	\\
14 & 	 23.1	&   21.2 & 	 20.5	&     20.4 & E	\\
15 & 	 21.2	&   20.1 & 	 19.6	&     19.6	\\
16 & 	 21.4	&   20.5 & 	 20.0	&     19.8	\\
17 & 	 20.6	&   20.0 & 	 19.4	&     19.5	\\
18 & 	 21.1	&   19.4 & 	 18.5	&     18.6	\\
19 & 	 22.5	&   22.4 & 	 22.0	&     21.6	\\
20 & 	 17.6	&   16.4 & 	 15.7	&     15.8	\\
21 & 	 21.9	&   20.5 & 	 19.5	&     19.2 & E	\\
22 & 	 22.0	&   21.4 & 	 19.9	&     19.7 & E	\\
23 & 	 21.6	&   21.6 & 	 20.8	&     20.6 & E	\\
24 & 	 19.4	&   17.9 & 	 17.0	&     17.0	\\
25 & 	 20.7	&   19.7 & 	 19.1	&     19.2	\\
26 & 	 22.4	&   19.3 & 	 17.8	&     17.2	\\
27 & 	 21.9	&   21.2 & 	 20.4	&     20.0 & E	\\
28 & 	 20.3	&   18.6 & 	 17.7	&     17.5 & E	\\
29 & 	 19.3	&   17.3 & 	 16.2	&     16.0 & E	\\
30 & 	 19.0	&   17.8 & 	 17.2	&     17.1 & E	\\
31 & 	23.0	&   20.2 & 	 18.8	&     18.4	\\
32 & 	 21.3	&   20.2 & 	 19.6	&     19.3 & E	\\
33 & 	 23.4	&   22.0 & 	 20.8	&     20.1 & E	\\
34 & 	 23.8	&   22.3 & 	 21.0	&     20.1 & E	\\
35 & 	 22.8(*)&   22.8(*) & 	 21.0(*)&     21.6(*)	\\
36 & 	 23.4	&   23.3 & 	 22.5	&     ----	\\
37 &    22.7	&   21.6 &     21.1	&   20.5	\\
38 &    20.4	&   19.4 &     18.9	&   18.9	\\
\end{tabular}
\end{center}
\end{table}
}

\subsubsection{{\bf Galaxies in/behind the filament}}
\label{.further1}

We want to pay attention now to  objects 1, 2, either embedded in the
filament that joins NGC 7603 and NGC 7603B or behind the filament.
The two objects appear  point like in our deep image in the R band (see
above). The field of the filament was observed in the F606W filter  with the
Hubble Space Telescope. It does not cover the other three
narrow emission line galaxies, but we can examine how extended 
the objects are in the filament. Figure~6 shows this image. The field is centred 
on the filament
between NGC 7603 and NGC 7603B and clearly shows  the two objects within it.
Both of them appear as extended objects; this is specially clear for object \#1 
(the one
closer to NGC 7603). The figure also shows a contour plot of both objects
which confirms the visual impression of both as extended objects. 
The FWHMs of objects 1 and 2 are $\sim 0.3$ and $\sim 0.4$ arcseconds respectively, 
which is rather small 
to be measured in a ground-based telescope with seeing of 1 arcsecond, and
seems to indicate that they are extended rather point like objects.

The two {\bf objects} in the filament are apparently a little  deformed,
although the significance is not too high (the two lowest isocontours in Fig.~6 are $\sim 2\sigma $ and $\sim 3.5 \sigma$ respectively over  the
average flux in the region).  The tail of object \#1 in the
northern part is warped pointing towards NGC 7603; and object \#2  has some
faint apparent tail in the northern part but this tail is less significant.

With the R band imaging we have studied the possible variability of these
objects. In addition to the weakness of these objects, the presence of the
filament makes  the  estimation of the magnitudes more difficult. In order to
carry out the measurements, we first fit a two-dimensional surface to the
filament excluding the region with the two objects. We estimate the
uncertainties in the magnitude for these two objects as $\sim 0.2$ mag. We 
have calibrated with standard stars only the R image taken  on  2000 June 13,
but we have performed differential photometry  of the other images with eight
bright  stars in the field. According to the mentioned uncertainties, we conclude the
absence of variability above $0.3-0.4$ mag.

\section{{\bf Spectroscopy}}
\label{.spectra}

The QSO candidates are in general too faint for spectroscopy with a 4.2 m telescope.
In any case we used this telescope: 1) to corroborate and improve the
spectra of both objects ($z=0.24$ and $z=0.39$) in the filament; 2) to obtain the redshift and classifications of some
other objects in the halo of NGC 7603 (objects like
\#17, \#21, \#22 were interesting  because of the peculiar position that they occupy with respect the halo and filaments
of NGC7603; (Fig.~1 shows that these sources lie within the halo of NGC 7603); 
3) to observe AGN candidates which are not too faint.

Table \ref{spectra_table} summarizes the objects crossed by the three positions of
the slit and a summary of the analysis of the spectra. Only the intense lines were used to determine
the redshift. Figur~4 plots the positions of these
slits. The spectra of the filament is poor because the slit in position 1 does not exactly
crosses 
 the maximum flux region of the filament, and the width of the slit (1.2 arcsec) is
small compared with the slitwidth of 5 arcsec used in the observations taken with the NOT
and presented in Paper I ([OIII] detections reported in Fig. 2b of Paper I were spurious). 
The {\bf main spectral features}  of these objects {\bf corrected of redshift} and the new ones (objects \#21, \#22, \#23) are shown in 
Figure~7.
All of them are narrow emission line galaxies.

Table \ref{Tab:EWs} gives the values of the equivalent widths of the different lines.
Apart from the errors pointed in the table due to noise in the
spectra, these equivalent widths are subject to the possible errors in the subtraction 
of the sky emission(+filament in objects \#1, \#2). Although
the absolute values of EWs can only be 
taken as a rough approximation, the ratio of close lines is rather exact (because here
the uncertainties in the continuum cancel). Roughly, the error would be a factor $\sim 2$ for
the continuum in the worst of the cases (assuming the error in the subtraction of the
sky+filament is equal to its Poissonian noise), which means that in the worst of the
cases the error of EW is a factor two too.

\begin{table*}
\caption{Spectral analysis. The error in the redshift is $\approx \pm 0.002$. The codes for the last two columns
are: 0-absorption emission line galaxy; 1-narrow emission line galaxy; 2-star; 
3-contamination by the filament; 4-spectra
very noisy.}
\label{spectra_table}
\begin{center}
\begin{tabular}{llllll}
Slit pos. & Object  & Spectral features ($>3\sigma $) & Redshift  & Type & Notes\\
\hline
\\
1,3 & NGC 7603B & CaH\&K, MgI, NaI, etc. & 0.056 & 0 \\
1 &  \#1     & OII, H$\beta$, OIII, OI, H$\alpha$ & 0.394 & 1 \\
1 &  \#2     & OII, NeIII, H$\beta$, OIII, H$\alpha$ & 0.245  & 1 \\
1 & Filament & H$\beta $, MgI, NaI & 0.030 &  Abs. & 4, Paper I \\
2 & \#17     & H$\alpha$, H$\beta$  & 0 & 2 & 3 \\
2 & 3.6"-WNW of \#17   &  ---- & ---& --- & 4 \\
2 & \#19   &  ---- & ---& --- & 4 \\
2,3 & \#21  & OIII$_{5007}$, H$\alpha$, NII, SII & 0.117 & 1 \\
2,3 & \#22 &  OII, H$\beta $, H$\alpha $ & 0.401  & 1 &  \\
3 & \#23 & H$\alpha $ & 0.246  & 1 \\
\end{tabular}
\end{center}
\end{table*}

\begin{table*}
\caption{Equivalent widths (in angstroms) of the emission lines of the five observed
narrow emission line galaxies. Errors include a rough
determination of the noise and the error in the determination of the continuum, but do not
include the error in the subtraction of the sky(+filament in objects \#1, \#2).}
\label{Tab:EWs}
\begin{center}
\begin{tabular}{llllll}
Line & 				Object \#1  & Object \#2 & Object \#21 & Object \#22  & Object \#23 \\
\hline \\
OII$_{3727}$ &			56.3$\pm$6.1	&36.2$\pm$7.3 &  37.1$\pm $9.3 &    15.3$\pm$2.1      & 68.4$\pm$75.1	    \\         
NeIII$_{3869}$ &				&13.0$\pm$3.1 &  	    &		    &		   \\	       
H$\beta $ 	&		13.5$\pm$1.1	&43.3$\pm$4.5 &  3.3$\pm $0.8   & 6.9$\pm$1.1	 & 3.7$\pm$1.5  	\\	       
OIII$_{4959}$ 	&		11.2$\pm$1.2	&62.8$\pm$6.3 &  	    &		    &		   \\	       
OIII$_{5007}$ 	&		19.2$\pm$1.7	&172.3$\pm$22.1&  2.6$\pm$0.5	  & 2.3$\pm $0.9    & 3.5$\pm$1.5	      \\	       
OI$_{6300}$ 	&		11.7$\pm$2.4	&	     &  	    &	   17.0$\pm $6.1	    &		   \\	       
H$\alpha $+NII$_{6548}$ &	81.0$\pm$21.9	&160.9$\pm$26.2&   15.6$\pm$0.7	    & 25.9$\pm $5.4	& 18.5$\pm $ 4.3    \\         
NII$_{6584}$ 	&		16.2$\pm$7.3	&8.8$\pm$4.6     &   8.4$\pm $0.5	 & 6.2$\pm$ 3.1  & 1.8$\pm $1.8	 \\	       
SII$_{6717+6731}$ &				&	     &    7.3$\pm$ 0.7   &		 &		\\	
\hline \\ 
$\log {\rm \frac{OIII_{5007}}{H\beta }}$ & 0.15$\pm$0.05  &  0.60$\pm$0.07 & -0.10$\pm$0.13 &  -0.48$\pm$0.18 & -0.02$\pm$0.26 \\
$\log {\rm \frac{NII_{6584}}{H\alpha }}$ & -0.70$\pm$0.23 &  -1.26$\pm$0.24 & -0.27$\pm$0.03 & -0.62$\pm$0.24 & -1.01$\pm$0.45 \\
\end{tabular}
\end{center}
\end{table*}

The spectral classification of narrow emission line galaxies is usually made
through the flux ratios of specific 
lines corrected of reddening, assuming a constant intrinsic ratio for $\frac{F_{\rm intr.}(H\alpha )}
{F_{\rm intr.}(H\beta )}$. However, this ratio changes when the physical situation in the
galaxies does not obey a simple model moreover, we have extinction from the dust in the own galaxy
plus the extinction of the filament in cases of \#1 and \#2, and some minor contribution 
from the Galactic
extinction, which are difficult to separate. In order to classify
the galaxies, we use the ratios of lines that are close in wavelength. The difference between the
flux ratio and the equivalent width ratio is neglected.
The spectral classification criteria is based on the ratios 
$\log {\rm \frac{OIII_{5007}}{H\beta }}$ and 
$\log {\rm \frac{NII_{6584}}{H\alpha }}$ (given in Table \ref{Tab:EWs})
(Veillux \& Osterbrock 1987, Filippenko \& Terlevich 1992,
Dessauges-Zavadsky et al. 2000), and gives the result that the objects in
table \ref{Tab:EWs} are HII-galaxies except object \#21, which might be either a HII-galaxy or a LINER (since its 
continuum is strong and the emission lines are faint, 
it might be a ``Low Luminosity Active Galactic Nuclei''; Maoz et al. 1998).

\subsection{{\bf Galaxies in/behind the filament}}
\label{.further2}

The HII-galaxies embedded in the filament, \#1 and \#2, seem to be indeed a quite peculiar 
star-forming galaxies. The very intensive $H\alpha $
(equivalent width: EW(H$\alpha )\approx 80$ \AA \ and 160 \AA \ resp.), 
if correct (i.e. if the continuum is really as low as obtained by us
and the sky+filament subtraction has not changed the level of the continuum; roughly, error of EW should be
a factor two at most), would be indicative from a vigorously star-formation galaxy. 
Only $\sim 2$\% and $\sim 1$ \% resp. of the normal HII-galaxies have a so high EW(H$\alpha$) 
(Carter et al. 2001).
However, if they were dwarf HII-galaxies, these high EWs would be within the normal expected values.
The mean intrinsic colour of these objects is $(B-V)_0$=0.22 mag and $(B-V)_0=0.10\pm 0.2$ resp. (with a dispersion of 
$\approx 0.10$ mag, plus an error of $\sim 0.15$ mag due to the factor 2-error in the value of EW; 
total: $\sim \pm 0.2$ mag) (Kennicutt et al. 1994, Fig. 2a).

We apply the correction of extinction for the flux of these objects in the following way. 
First, we derive the observed $(B-V)_0$ in the reference system 
of the galaxy [i.e. we calculate the equivalent (B-V) in the redshifted
wavelengths; we do this through the calculation of the flux in the corresponding wavelengths of the
redshifted B and V filters given the UBVRI fluxes; this is equivalent to make the k-correction]. 
These are: $(B-V)_0=0.58$ (object \#1)
and $(B-V)_0=0.45$ (object \#2). We neglect the difference between the colours of a face-on galaxy
and other inclinations. Therefore, the differences between these measured colours and the colours expected
for these HII-galaxies with the corresponding $EW(H\alpha +[NII])$ are: 
$\Delta (B-V)_0=0.48\pm 0.2$ (object \#1) and $\Delta (B-V)_0=0.23\pm 0.2$ (object \#2). 
We assume that the measure of the colour has negligible error (the absolute magnitude in each filter
has 0.2 mag. of error, due to the contamination of the filament, but in the measure of the colour, 
since the technique to decontaminate from the filament is the same, this error cancels). 
We attribute these differences to the extinction produced by the filament 
($z_{\rm fil}=0.030$) plus the Galactic extinction ($z=0$):

\[
\Delta (B-V)_0=A_{\rm Gal.}[\lambda _B(1+z)]-A_{\rm Gal.}[\lambda _V(1+z)]
\]\begin{equation}+
A_{\rm fil.}\left[\lambda _B\frac{(1+z)}{(1+z_{\rm fil })}\right]-A_{\rm fil.}
\left[\lambda _V\frac{(1+z)}{(1+z_{\rm fil })}\right]
.\end{equation}
The Galactic reddening is relatively 
low: [$A_{\rm Gal.}[\lambda _B(1+z)]-A_{\rm Gal.}[\lambda _V(1+z)]=0.03$  
for both objects according to Schlegel et al. (1998) maps of extinction.
Assuming a reddening due to the filament similar to the Galactic dust [$A(\lambda )/A_V$ from
Mathis 1990], we get

\begin{equation}
A_{\rm fil.(obj.\ \#1)}[\lambda '=\lambda /(1+z_{\rm fil})]=(2.0\pm 0.9)\frac{A(\lambda ')}{A(\lambda _V')}
\label{extinc1}
,\end{equation}
and
\begin{equation}
A_{\rm fil.(obj.\ \#2)}[\lambda '=\lambda /(1+z_{\rm fil})]=(0.9\pm 0.9)\frac{A(\lambda ')}{A(\lambda _V')}
\label{extinc2}
,\end{equation}

The corrected magnitudes, including Galactic and filament extinction correction, 
are written in Table \ref{var_table}. 

The calculation of the absolute magnitudes, for instance for the filter $V$ [in the own galaxy, i.e.
$\lambda =\lambda _V(1+z)$], can be carried out by means of:

\begin{equation}
M_V=m_{\rm corr.ext.}[\lambda _V(1+z)]-5\log (D_{\rm Mpc})-25
,\end{equation}
\[
D_{\rm Mpc}=\frac{c(1+z)}{H_0}
\]\begin{equation}\times
\int _0^z [(1+x)^2(1+\Omega _Mx)-x(2+x)\Omega _\Lambda ]^{-1/2}dx
.\end{equation}
We assume $H_0=71$ km/s/Mpc, $\Omega _M=0.27$, $\Omega _\Lambda =0.73$
(derived from WMAP data; Bennett et al. 2003).
If we consider the redshifts as indicator of the cosmological distance: $M_V(object \#1)=-21.5\pm
0.8$,  $M_V(object \#2)=-18.9\pm 0.8$. If we considered an anomalous intrinsic redshift case  (in such
a case, in order to derive the distance, we set $z=z_{\rm fil}=0.03$), the results are: $M_V(object
\#1)=-15.2\pm 0.8$, $M_V(object \#2)=-13.9\pm 0.8$. In this second case, they would be in the faint
tail of the HII-galaxies, type II (Telles \& Terlevich 1995); they would be dwarf galaxies, ``tidal
dwarfs'' as in Stephan's Quintet (Lisenfeld 2002) but with anomalous redshifts, and this would explain
the observed strong star formation ratio: objects with low luminosity have higher EW(H$\alpha $)
(Carter et al. 2001). Teplitz et al. (2003)  show examples of compact emission line galaxies with
very  high EW(H$\alpha $) and absolute filterless  magnitudes between -14 and -15 (e. g., SPS
J082344.12+292351.3).

\subsubsection{Comparison with Paper I and corrections}

With these new spectra, we confirm the redshifts of objects \#1 and \#2
observed in Paper I, and furthermore we detect
the H$\alpha$ emission line in the spectra of both objects. Now it is
possible a more accurate estimation of the linewidth of each object.

In Paper I, the tenptative possible classification of both objects 
as QSOs (indeed, in Paper I we claimed that they were compact emission line objects, either
QSOs or HII-galaxies) is not confirmed here.
In Paper I, we had not analysed HST data so we could not see whether the objects had
any extension. 
In Fig. 3 of Paper I, we  pointed out that the $H\beta $ line in  object \#2 had
a FWHM of 49 \AA \ while the forbidden lines had 30 \AA \ in these low resolution spectra;
and the same thing was observed for object \#1 with poorer signal/noise. 
However, in the present WHT-telescope higher resolution spectra, we have not observed
this relative broadening, so we must think that the apparent excess broadening of 
H$\beta $ in Paper I was an artefact due to noise.

In Paper I, we made a rought measurement of  the parameter 
$R_{23}$ directly from ratios of equivalent widths\footnote{There 
was also an erratum in the reference to Filippenko \& Terlevich (1992) in Paper I: 
they said that LINERs have $\log _{10} \frac{OIII _{5007}}{H\beta }\le 0.5$ instead
of $\frac{OIII _{5007}}{H\beta }>0.5$, and this is used to separate Seyfert 2s and LINERS
rather than HII-galaxies from LINERs.}; however, lines fluxes corrected
for reddening would be necessary. The results presented in this paper by making
use of EWs of close lines give in any case the same conclusion: they are HII galaxies
(provided that they have narrow lines).

The slight differences in the R-magnitude (0.2 and 0.1 mag
respectively) with respect to the values presented in Paper I are caused by 
differences in which the filament was subtracted and are within the error.
In Paper I, it was claimed that the objects have $m_{b_j}<21.9$ (corrected of extinction), 
and this is correct, but for
a reason different from the arguments given in Paper I. They are intrinsically blue,
but because of the extinction they are observed as red
[$(B-V)_0=0.10$ and 0.22 respectively for objects \#1 and \#2 corrected for 
extinction and the k-correction] instead of blue as claimed in Paper I.

Summing up, we confirm the main results of Paper I, except in the tenptative statements
such as the possible classification of objects \#1 and \#2 as QSOs.

\section{Analysis}
\label{.analysis}

\subsection{A candidate perturber}

According to the results in \S \ref{.diffuse}, some 
close and not very faint neighbour must be in the surroundings of NGC 7603.
If we assumed that NGC 7603B and NGC 7603 have the same distance, this filament
would be clearly due to the interaction among both. Are there other possible candidates?

There is a galaxy with similar redshift, one magnitude fainter, and 10.3 arcminutes 
from NGC 7603: NGC 7589; or B231533.01-000313.1, three magnitudes fainter and
12.6 arcminutes of distance. However, both of them are in the opposite direction of the filament
(in the west instead of the east). 
We do not find any other appropriate candidate for the interaction in the 
surrounding 30 arcminutes. Nonetheless, we cannot be sure that this companion object
does not exist until we make the spectroscopy of all the surrounding objects around
NGC 7603. For instance, galaxy \#29 (see Fig.~4 and table \ref{ugri_table})
is four magnitudes fainter than NGC 7603 in B-filter, it has an angular distance of 2.5 arcminutes
(linear distance larger than 100 kpc) and it seems to be in the direction of the tail which is extended
towards north; it might be a candidate of companion to produce the tidal disturbances. Or galaxies \#28, \#30.
Johnston et al. (2001) in their eq. (11)/Fig. 6 calculate the expected surface brightness
magnitude in such cases. Assuming $t>1$ Gyr 
a mass-to-light ratio of 10, rotation velocity from NGC 7603 of 200 km/s in the outer disc,
and a distance of the satellite of 100 kpc we would have that the observed surface brightness in R of the
filament should be $>27$ mag/arcsec$^2$; however we observe that it is 24 mag/arcsec$^2$.
Therefore, it seems in principle not too probable, but it might be, and this remains as a possible
solution within a standard cosmological redshift hypothesis scenario until it can be discarded
with new spectroscopical observations of all the surrounding objects and/or calculations
which proof that these faint objects cannot produce the observed tidal disturbances.

\subsection{Statistics}

In this section, we will carry out some statistical calculations in order to show
how anomalous is the observed configurations: 1) to have up to two QSOs in the neighbourhood
of NGC 7603; 2) to have the observed configuration of 4 objects with different redshift
connected by a filament.

\subsubsection{QSOs statistics}

In \S \ref{.colours}, we  concluded that there are three objects that follow 
eq. (\ref{selecQSO}) within a radius 
$R=1.5$ arcminutes from the center of NGC 7603 (the distance to object \#36). 
One of these (object \#23) is not a QSO (see \S \ref{.spectra}). 
The other two were too faint to be observed spectroscopically
 with the available telescope and time (perhaps a larger telescope
would be needed).
Therefore, we have at most two QSOs in the field of NGC 7603 [indeed, some
extra QSOs are possible, but with a low probability because
the multicolour criterion covers 80-90\% of all QSOs (see \S \ref{.colours})]
with $m_{b_j}=22.6$, 23.6 (respectively for objects \#19 and \#36; derived from Sloan filters
information as in \S \ref{.further2}). 

The probability for such a event is (assuming roughly that $p_i$, the
probability for the detection of each QSO, follows $e^{p_i}\sim 1$):

\begin{equation}
P_Q\ge \frac{(\pi R^2)^2N_{\rm QSO}(m_{b_j}<22.6)N_{\rm QSO}(m_{b_j}<23.6)}
{2!}
.\end{equation} 
According to the QSO counts from eq. (\ref{nbj}), the probability 
is $P_Q \ge 0.029$, so one might expect these number of 
background QSOs to occur by chance (2.2-$\sigma $ at most, if both candidates are
confirmed as QSOs).

\subsubsection{Probability of NGC 7603 and its 3 companions being
a chance projection effect}
\label{.probab}
 
From Figs. 1 of Paper I; and Fig.~4 or 6 of this paper, it seems extremely improbable that four objects at 
different distances
can show a chance projection in the way these figures reveal.
Statistics have been calculated in several ways for some time concerning 
the anomalous redshift problem (e.g., Arp 1981, 1999$a$; Burbidge et al. 1997), in order 
to assess the probabilities
of peculiar configurations. However, many
other authors (e.g., Noerdlinger 1975; Sluse et al. 2003) have suspected that many of 
these calculations are unappropriate. Some authors also say that
one should not carry out a calculation of the probability (``a posteriori probability'') 
for an a priori known
configuration of objects (for instance, that they are aligned, or that they
form a certain geometrical figure) because
all possible configurations are peculiar and unique.
For example, if the Orion constellation is observed and we want 
to calculate the probability of their stars to be projected
in that exact configuration, we will find that the probability is nil
(it trends to zero as the allowed error in the positions of the stars
with respect the given configuration goes to zero),
but the calculation of this probability is worthless because
we have selected a particular configuration which has been  observed a priori.
Therefore, the statistics to be carried out should not concern the geometrical 
figure drawn by the sources, unless
that geometrical configuration be representative of a physical process
in an alternative theory 
(for instance, aligned sources could be representative of the ejection
of sources by a parent source). In this last sense, we think that
many of the statistics already published are worthwile and indicate the reality of some statistical anomaly.
The real question is to look for peculiarities associated with
peculiar physical representations, not just peculiarities in the
sense of being unique.

For our case, we will use a simple fact: the connection of four objects
throughout a filament. This aspect represents a physical peculiarity, not because of
their uniqueness but because they could be better represented  by an
alternative theory claiming that the four galaxies are at the same distance,
three of them ejected with the filament by the parent galaxy
NGC 7603. We are not going to determine the  probabilities
of forming triangles or any shape observed a priori. The peculiarity
that we want to analyse is not comparable with the previous example of
Orion because we have in mind a physical representation rather than
the given peculiar distribution of sources.
The difference from the Orion problem 
is that the peculiarity of Orion is not associated
with any particular physical representation to be explained by an alternative theory.
The question is as follows: what is the probability, $P$, of the
apparent fact arising from a random projection of sources with different distances?
Or, in other words, what is the probability, $P$, that the standard theory can explain the
observed facts without aiming alternative scenarios?
This probability is as follows: NGC 7603 has a filament of area $A$.
The probability of having three further independent sources, with the
corresponding magnitudes of the objects 1-3, projected on that filament is
(assuming that the individual probabilities for each event $p_i$ follow 
$e^{p_i}\approx 1$):

\begin{equation}
P_=\frac{A^3N_1(m\le m_1)N_2(m\le m_2)N_3(m\le m_3)}{3!}
\label{p1}
,\end{equation}
where $N_i$ is the source density on the sky for the type of sources
of  object $i$ with apparent magnitude less than $m_i$
(magnitudes corrected for Galactic+filament extinction, in order to be comparable
with the galaxy counts from appendix \ref{.counts}), for the filter
in which we know the magnitude of the source. We will use, for instance,
filter B, but the statistics would give similar results for any filter.
Some authors (e.g., Sluse et al. 2003, hypotheses H2-H3) use in the calculation of the probabilities
the limiting magnitude of the survey instead of the magnitude of the object, 
which gives a higher probability. However, this is not totally correct 
because, randomly, one would expect  most of the detected objects
to be close to the limiting magnitude. If this method is followed, the magnitude
of the object and the limiting magnitude of the survey are very close and
there are no major differences in the calculation; but, in the case the magnitude
of the object being much brighter than the limiting magnitude one should multiply
$P$ by a factor that characterizes the probability  this object
being much brighter than the limiting magnitude (the brighter it is, the lower
the probability), and this is equivalent to use the magnitude of the
object. So we think that Sluse et al. (2003) hypotheses H2-H3 are inappropriate.

The area of the filament is approximately 35 arcsec in length multiplied by 
4 arcsec in width (the area plotted in Fig.~6):

\begin{equation}
A\approx 35"\times 4"=140\ {\rm arcsec}^2=1.1\times 10^{-5}\ {\rm deg}^2 
.\end{equation}

We are not going to use other peculiarities of the system like
the fact that  objects \#1 and \#2 are
positioned where the filament contacts  NGC 7603B and NGC 7603 respectively.
Neither,  are we going to use the fact that two of the three galaxies are
HII-galaxies; we  pay no attention to galaxy type.
Neither,  are we going to use the distribution of redshifts 
(from major to minor). 
If we took these facts into accounts, the probability $P$ would be
somewhat lower. 

NGC 7603B is a galaxy with $m_{B,1}=16.6$ (Sharp 1986, corrected only for
Galactic extinction; it would be less if the foreground filament produced any extinction
in the galaxy). And the magnitudes corrected for extinction
of the two HII-galaxies are: $m_{B,2}=21.1\pm 1.1$ and $m_{B,3}=22.1\pm 1.1$.
With all these numbers, and the counts given by eq. (\ref{countsgalb}), 
the deduced probability is

\begin{equation}
\log P=-8.6\pm 0.8
.\end{equation}
The error is large, due to the uncertainty of 1.2 mag in the objects \#1 and \#2, but the order of magnitude
does not change too much.
This means that we have a probability of a few times $10^{-9}$ of finding 
three galaxies of any type by chance with different distances projected on
a filament (an arm or another structure) with an area of 140 arcsecond$^2$ of an arbitrary galaxy
with respective apparent magnitude (corrected for extinction) 
less than or equal  16.6, 21.1, 22.1 respectively, and somewhat 
higher if the magnitudes are up to 1.2 fainter in the last two objects.
If there were no filament extinction at all, the value of $\log P$ would 
be -7.1.
There are certain facts that could make the probabilities calculated above
larger. For instance, the density of any of the objects would be 
significantly larger than  we have assumed, if the distribution of
any of these sources were clustered in some specific regions, or if there were bias
selection effects. There is not justification for  a conspiracy in which our line
of sight crosses three clusters of galaxies at different redshifts ($z=0.056$, $z=0.245$
and $z=0.394$). However, maybe at least one of the objects is in a cluster (for instance the object
at $z=0.245$ since we have found another object with $z=0.246$ not very far away).
In any case, the increase in the probability due to the increase of the density
in lines of sight with clusters is compensated with the additional factor to be 
multiplied
by the present amount $P$ to take into account the probability of finding clusters
in the line of sight. On average, in all the arbitrary lines of sight on the sky,
the probability will be given anyway by the above value of $P$ 
(see further discussion in \S \ref{.clusters}). With regard bias selection effects,
there are no such biases, because we have used complete galaxy counts from complete surveys up
to a given magnitude (appendix \ref{.counts}). 

Furthermore, we could multiply $P$ by the probability 
to have two extremely vigorous star formation
in the HII-galaxies, $P_2$. 
The calculation of $P_2$ (with probabilities $\sim 1$\% and $\sim 2$\% 
for each galaxy, as said in
\S \ref{.spectra}) is
\begin{equation}
P_2=0.01\times 0.02=2\times 10^{-4}
\end{equation}
There is nearly independence between both probabilities, so the global probability
is the product of $P$ and $P_2$. There is a correlation between absolute 
magnitudes and the
star formation ratio: the fainter the HII-galaxy, the higher is the star
formation ratio (Carter et al. 2001). As a matter of fact,
Teplitz et al. (2003) find in their sample that many galaxies 
have large values of EW(H$\alpha $) but their objects
are in general intrinsically fainter. However, as has already been said, 
if we accept the cosmological redshift hypothesis, 
our two HII-galaxies would have average or high luminosities (absolute
magnitudes in V from $-19$ to $-22$); therefore, it is not possible and enhancement of the probability due to 
some selection effect (neither Malmquist bias nor the opposite one).
In any case, we will not consider the low value of $P_2$ until we can have more accurate measures
of EWs, and we will only discuss the probability $P$ quoted above. 

We have found this extraneous combination of circumstances (if we adopt the
hypothesis of cosmological redshifts) by accident. We did not sistematically analyze all
the 
Seyfert galaxies like NGC 7603 in order to find something like this. Even if we
have made a complete analysis of these characteristics,  to observe only
one object with a probability $3\times 10^{-9}$ would  still be very small.
According to SIMBAD there are 237 AGN-galaxies in all the sky with a B magnitude 
less than 14.0 (the magnitude of
NGC 7603; de Vaucouleurs et al. 1991). Therefore, the probability to have an AGN
with a B magnitude  up to 14.0 with a cluster of coincidences that we observe in NGC 7603 is:

\begin{equation}
P_{\rm all\ AGNs, m_B<14.0}\sim 7\times 10^{-7}
.\end{equation}
Some alternative theories (Arp \& Russell 2001, Burbidge 1999b)
do predict that Seyferts eject galaxies
so the statistics with AGN makes sense.
However, even if we decided to use all the galaxies, independently of whether they are
AGNs or not, according to SIMBAD the number of galaxies with $m_B<14.0$ is 3655, so the
probability of finding a galaxy
with a B magnitude  up to 14.0 with the cluster of coincidences that we observe in 
NGC 7603 is:

\begin{equation}
P_{\rm all\ galaxies, m_B<14.0}\sim 10^{-5}
,\end{equation}

This assumes that all the galaxies have some filament/arm like NGC7603 to find  the
background objects, which is not necessarily the case, so we are overestimating the
probabilities. And these probabilities are calculated assuming that there is only one
case with  this cluster of chance circumstances. This could not be the case because not all the
systems have been studied in detail, and some of the cases which were studied with some
detail also present   similar coincidences. For instance, cases like 3C212 (Stockton
\& Ridgway 1998), NGC 3067 (Carilli et al. 1989; Carilli \& van Gorkom 1992), NGC 3628
(Arp et al. 2002); NGC 1232, NGC 4151 or NGC 622 (Arp 1987); or the cases mentioned 
in Arp (1980) also present 
some filaments/arms that have at their ends, or somewhat beyond the end in the 
direction
of the filament, some galaxy or QSO with different redshift (Note: many of 
these examples have however magnitude large than 14.0 in B). Another possible
example might be the tail  between NGC 7320 and NGC 7318 in Stephan's Quintet
(Moles et al. 1998; Guti\'errez et al. 2002; Williams et al. 2002). In some of these
tails the presence of dwarf galaxies looking like objects 1 and 2
has been discovered even at relatively large distances from the disrupted galaxy (Gallagher et al. 2001).  So,
again, we are overestimating the probability because we are only considering one case among all
galaxies, and there are other coincidences too.

It is remarkable that the presence of the filament gives the configuration a low
probability, but even without taking into account the presence of the filament, the
probability is still somewhat low too. Given a square of diagonal 59 arcseconds (the separation
between NGC 7603 and NGC 7603B), the chance to have there four galaxies with magnitudes
in the B-filter of 13.8 (the magnitude of NGC 7603 corrected for  extinction), 16.6, 21.1 and
22.1 would be $3.5\times 10^{-11}$ [again with  Poissonian  statistics and the 
counts given by eq. (\ref{nbj})]. There are $3.2\times 10^8$ squares like this on the whole
sky, so the probability of finding only one square in the whole sky with this congregation of
four objects of different distances is $1.1\times 10^{-2}$. A 1\% probability is not
extremely low, but taking into account that this is the probability for the whole sky,
and that it would have average properties, this low number is also somewhat strange.

Summing up, even in the worst of the cases, we have that the probability to find only one case like
the present three background galaxies of the given magnitudes 
in the filament/arm among all galaxies up to magnitude
14.0 in B-filter is around $10^{-5}$, so we
consider that it is justified to talk about ``anomaly''.
The question now is to find the reason for this low probability and whether
it can be explained in terms of a cosmological redshift or not.

\subsection{{\bf Possible explanations for the low probability observed configuration}}
\label{.discussion}

Possible scenarios to explain the present case of NGC 7603 depend on
the possible explanations for the redshift of the objects are (Narlikar 1989;
Hoyle \& Burbidge 1996): cosmological (with the observed configuration
due to clusters in the line of sight, or gravitational lensing), Doppler, gravitational or others.
In this section, we are going to discuss how well the different hypotheses
 explain the present case.

\subsubsection{Clusters in the line of sight?}
\label{.clusters}

Perhaps we have found a line of sight with many clusters of galaxies, that increases
significantly the density of sources with respect to a Poissonian distribution. However, 
as explained in \S \ref{.probab},  a conspiracy in which our line of sight crosses three
clusters of galaxies at different redshifts ($z=0.056$, $z=0.245$ and $z=0.394$) is not
justified because the increase in the probability due to the increase of the density in
lines of sight with clusters is compensated with the additional factor to be multiplied
by the present amount $P$ to take into account the probability of finding clusters in the
line of sight. In average, in all the arbitrary lines of sight of the sky, the
probability will anyway be given  by the above value of $P$. We can illustrate this
argument with some simplistic calculations. Let assume that the clusters in the sky have
the same size, $A_c$,  a Poissonian distribution, and the same number of galaxies up to a
given magnitude, $n_c$ (galaxies/cluster).  This is a very rough model, because it is
clear that $A_c$ and $n_c$ depend on the redshift;  however, for our present
 arguments, the estimation with mean values of  $A_c$ and $n_c$ is enough. In such a
case, the total counts of galaxies, $N$, is:

\begin{equation}
N=N_f+N_cn_c 
,\end{equation}
where $N_f$ is the density of field galaxies (galaxies/deg$^2$) and $N_c$
is the density of clusters (clusters/deg$^2$). 
An example of a probability calculation   would be the one to have
three galaxies belonging to three different clusters in the area
$A$ (we assume that they have the same magnitude, for a simplistic calculation,
although it can be generalized to any magnitude distribution), i.e. 
the probability  three clusters being  in the
line of sight multiplied by the probability of three galaxies 
from different clusters being in the area $A$ of the filament given the density
of galaxies in a cluster,

\[
P=\frac{(N_cA_c)^3}{3!}\frac{\left(A\frac{n_c}{A_c}\right)^3}{3!}
\]\begin{equation}=
\frac{A^3N^3}{3!}\left(1-\frac{N_f}{N}\right)^3\frac{1}{3!}<\frac{A^3N^3}{3!}
,\end{equation}
that is, the probability is lower than $\frac{A^3N^3}{3!}$, which is
the probability we calculated in (\ref{p1}). Therefore, the supposition
of a line of sight with three clusters in the line of sight would make
the probability smaller instead of larger, and similarly for a lower
number of clusters.

Indeed, it is not likely to find clusters of galaxies at $z=0.245$ and $z=0.391$, in spite
of the two pairs of HII-galaxies with close redshifts,
because HII-galaxies are much less common in clusters than in field galaxies
(Gisler 1978; Dressler et al. 1985; Biviano et al. 1997), i.e. the probability
of finding HII-galaxies in clusters is much lower than among field galaxies.
Therefore, unless we want to introduce a new factor that further  reduces 
the value of $P$ (which makes the problem more difficult to solve in terms
of cosmological redshift rather than solving it), we must not see the solution 
of the clusters as
a way to explain the problem in normal terms of probability .

Nonetheless, although the low probabilities cannot be justified by this
scenario of clusters, and although the high star formation ratios seem
to point in the opposite direction, we also have  object \#23 with nearly
the same redshift as object \#2, and object \#22 with a difference of 0.007
in redshift with respect to object \#1. They are not necessarily in the same
cluster, but perhaps they form small groups of galaxies with separations 
of 0.5 or 2 Mpc (for the pairs at $z=0.25$ and $z=0.40$ respectively).
Therefore, this standard scenario, even though it cannot explain the statistical
problem, should be still borne in mind. We have considered above 
that the distribution of clusters is Poissonian;
it might be that we have detected two or 
three clusters in the line of sight for some special reason.
Could our line of sight be tangential to a wall or sheet within the 
large scale structure, for instance? This seems difficult to imagine, since
we would need a wall of size 2 Gpc. The Hydro-Gravitational Theory
(Gibson 1996; Gibson \& Schild 2003) would claim
that the members of a cluster (NGC 7603, NGC 7603B, object \#1 and object \#2)
formed together, and that they remained together until 
the uniform expansion of space in the universe finally overcame the gravitational
 and frictional forces of the cluster, and the different galaxies separated
with very small transverse velocity with respect the line of sight because of 
the halo gas friction and their sticky beginning. 
The stretching would be along a pencil beam of length $\sim 2$ Gpc in the line
of sight by the expansion of the universe, but  a 
preferred direction of the expansion instead of an isotropic expansion it is 
not justified; 
why the expansion along the line of sight is not sticky as it is for 
the other directions? These are in any case speculative possibilities which
are not compatible with the CDM theory of the formation of the 
large scale structure. The question, therefore, remains open.

\subsubsection{Gravitational lensing}

A better explanation might  in principle be found if we considered some kind of gravitational
lensing. For instance, amplifications up to a factor $\sim 30$ are expected
(Ellis et al. 2001) for background objects apparently close to the central parts
of massive clusters. The effect produced by an individual galaxy like NGC 7603 should be much
smaller, and the low redshift galaxy ($z=0.029$) NGC 7603, 
as the putative lens of very distant sources ($z=0.245$ and $z=0.394$)
would have a very small amplification because of the large angular distance of the
sources. Some rough calculations can illustrate this argument;
given a galaxy with Einstein radius $\theta _E$, the enhancement
in the density of background objects as a function to the angular distance,
$\theta $, to this galaxy will be (Wu 1996):

\begin{equation}
q_Q(\theta )=\frac{N[m<m_{b,lim}+2.5\log \mu (\theta )]}{N(m<m_{b,lim})}\frac{1}{\mu (\theta)}
\label{gravlens}
,\end{equation}
where $\mu $ is the magnification factor (Wu 1996),
\begin{equation}
\mu (\theta )\approx \frac{\theta }{\theta -\theta _E}
.\end{equation}
In order to have a value of $P$ that is not very low, we would need this to be
$\sim 10^3-10^6$ higher, i.e. an average enhancement of $\sim 10-100$ in
density for each of the galaxies. With the counts of eq. (\ref{countsgalb}),
for the lowest enhancement, this requires an average magnification of $\mu
(\theta )$ of $\sim 2\times 10^4$. We would need to be in the ring  $\theta
_E<\theta <\theta _E (1+5\times 10^{-5})$, which is very narrow  with a
very low probability; so again the problem is not solved by this artefact.  It
is clear from eq. (\ref{gravlens}) that the density of sources does not
increase so quickly, unless the counts increase extremely quickly with the
limiting magnitude, which is not our case. This is so because the enhancement
in the source counts increases because of the flux increase of each source but
decreases because of the area distortion, which reduces the number counts by losing
the sources within a given area (Wu 1996).

In our case, since the distance of the sources to the centre of NGC 7603 is
0.5-1 arcminute, we would need either a very large value of the Einstein radius
of the gravitational lens placed in the centre of NGC 7603, which would require
a huge mass (for instance, an average E/S0 galaxy has a $\theta _E\approx 1.33$
arcseconds, Wu 1996), or that the gravitational lenses be not so massive but much
closer to the magnified objects. The first possibility may be automatically
rejected, since even in the case that NGC 7603 had the mass of a cluster of
galaxies, the magnification would affect at most only one of the three
objects in the filament, the one closer to its Einstein radius. The second
hypothesis, the possibility that multiple minilenses are distributed in the
halo of the galaxy, has already been proposed: gravitational mesolensing by King
objects (Baryshev \& Bukhmastova 1997; Bukhmastova 2003). The strong
gravitational lensing would be produced by King lenses: globular clusters
(Bukhmastova 2003), dwarf galaxies, or clusters of hidden mass with masses
between 10$^3$ and 10$^9$ M$_\odot $. This is an interesting idea, although we
are not convinced by the proof presented by one of authors of the idea
(Bukhmastova 2001) which reveals excesses of pairs of galaxy/QSO with
$z_{gal}>0.9z_{QSO}$, because many of these pairs were indeed the same object
classified both as QSOs and galaxies. Anyway, the idea is interesting, and it
might be considered as a serious proposal for solving the statistical correlations
between QSOs and galaxies in large surveys. Nevertheless, in our particular
case, it does not solve the low probability $P$,  because, as has been said, 
only in
narrow rings is the enhancement high enough, and these narrow rings have a very
small area, so, again, the probability of these being a large number of sources is
small.

\subsubsection{Non-cosmological redshift hypotheses}

The relative angular configuration of NGC 7603, NGC 7603B, object \#1 and \#2,
the filament connecting all of them (and, the probability that two of two
HII-galaxies in the filament have very high star formation ratios, if we
accepted as valid the measures of the EWs) would be naturally explained as a
consequence of a physical interaction between them. An interpretation which
explains the configuration as equivalent to other systems in interaction would
be clearly preferred over one in which the configuration is purely a
projection effect according to the calculations in \S \ref{.probab}.  In that
case, the filament would be a sign of disruption in NGC 7603 owing to the
proximity of NGC 7603B. This is reinforced by the fact that both NGC 7603 and 
NGC 7603B show asymmetries in the halo. The narrow emission line galaxies 
\#21, \#22, \#23 in the other side of NGC 7603 might also be embedded into the
extension of the halo pointing to these objects. 

In such a case, the redshifts would be non cosmological.
Some of the possible explanations for an intrinsic redshift 
with standard physics are now discussed:

\begin{description}

\item [Doppler:] 
Considering only the system of NGC 7603 and NGC 7603B, which have a difference
of around 8000 km/s is it possible that both galaxies are
at the same cosmological distance and that the difference in redshift reflects a
difference in peculiar velocities?  The known examples of interacting galaxies 
in the field show differences in velocity between  to $\sim 1000$
km$\,$s$^{-1}$. The larger density of objects in a cluster of galaxies and
the dispersion of velocities within them could favour high speed collisions
with differences  in velocity between the interacting galaxies of a few $\sim
1000$ km$\,$s$^{-1}$. The possibility of an encounter between groups of galaxies
with a difference in velocity of $\sim 4000$ km$\cdot$s$^{-1}$ has been considered
by de Ruiter et al. (1998) as a possible explanation of the peculiar field
around B2 1637+29. However, as far as we know, no example of such a
collision  in the field with a difference in velocity as large as that
existing between NGC 7603 and NGC 7603B has been reported so far, and it would be
unexplained in the framework of models of galaxy formation.
Furthermore, the extremely high velocity differences of the HII-galaxies would disrupt the
system quickly and there would be cases of blueshifts (in this or other anomalous redshift cases).

\item [Gravitational:] 
Anomalous redshifts could alternatively be explained in terms of
highly collapsed matter (Narlikar 1989).
The gravitational redshift explanation could then be an explanation
in terms of standard physics although we would need either very high masses
or very low radii for these objects. High mass seems to be excluded
since this would affect the rotation curves in the QSO-galaxy pairs
(Hoyle \& Burbidge 1996). Very dense non-high mass objects could explain the
situation, but the HII-galaxies and NGC 7603B are extended
objects; unless most of the mass is concentrated in the very centre of the
nucleus, giving an intrinsic redshift, and the outer part of the galaxies
had normal cosmological redshifts. At present, we have not detected these
differences of redshifts within the HII-galaxies, and it can surely discarded
for NGC 7603B.

\item [Multiple scattering:]
Dynamic multiple scattering has been also proposed to
explain these systems. Results in statistical 
optics (Wolf 1986; Datta, Roy \& Moles 1998$a,b$) show that a shift
in the frequency of spectral lines is produced with redshift independent of the
frequency when the light
passes through a turbulent (or inhomogeneous) medium, because of multiple 
scattering effects (Roy et al. 2000).
Perhaps, the anomalies could be caused by certain special conditions in the surroundings of the anomalous
redshift objects. 
Indeed, the scattering solution was proposed a long time ago as
a way to explain the loss of energy of the photons (``the tired light theory''), an
alternative to the cosmological redshift.
There were several proposals in terms of photon-photon or photon-matter
interaction due to some quantum effects (e.g., Finley-Freundliech 1954; 
Pecker et al. 1972;
Laio et al. 1997;  Moret-Bailly 2001). Potentially, this effect could explain the 
high redshifts of some anomalous redshift objects, since light travelling through their outer
atmospheres could be redshifted before leaving it, and the blurring would not be a problem 
here since the distance travelled is short.

\end{description}

\subsubsection{Variable mass hypothesis}

Apart from the mechanisms given  in the last subsection, non-standard physics 
has  also been used
to explain the redshift problem.
Hoyle \& Narlikar (1964) developed a new theory of gravitation
with particle masses depending on time according to $m\sim t^2$ and redshifts
\begin{equation}
1+z=\frac{\lambda _{source}}{\lambda _{observer}}=
\frac{m_{observer}}{m_{source}}
,\end{equation}
where ``observer'' and ``source'' stands for the measures from the different system at
the Earth and in the source respectively.
An explanation that these authors give for anomalous redshift galaxies is that new matter
is being created there with $t=0$, $m_{source}=m(t=0)$ for that new matter
and that the mass varies with the age (Narlikar 1977; Narlikar \& Arp 1993) to 
produce different redshifts.

\subsubsection{Higher redshift galaxies ejected by a parent galaxy?}
\label{.eject}

Some proposed models (e.g., Arp 1999a,b; Arp \& Russell 2001; Bell 2002a,b)
assume that some  QSOs are ejected by a parent galaxy and decrease in redshift
as they move outward, often although not always along the minor axis (the more
recent ejections are near the axes, but they later move away  because of
peculiar motions, precession of the galaxy or the spin axis of the nucleus; Arp
1999b), until they reach a maximum distance of $\sim 500$ kpc when they fall
back to the parent galaxy and turn into compact, active galaxies and, when they
are older, into normal galaxies. Galaxies would beget galaxies; they would not
be made from  initial density fluctuations in a Big Bang Universe (Burbidge
1999b). It is usually claimed that the variable mass hypothesis is the
explanation for the intrinsic redshifts. However, the scenario of ``galaxies
beget galaxies (with different redshift)'' should be considered as a separate
matter from the variable mass hypothesis or the Quasi Steady State Theory
because other explanations of the redshifts and other cosmological scenarios
could be compatible with the present idea.

This scenario seems to fit quite well the observed system: we would have three
(or possibly only two or one, if we considered that some of the galaxies are
background galaxies but not all of them) ejected together with the material of
the filament, and  we could think that any of the objects \#21, \#22 or \#23
might be part of the ejection on the other side of the galaxy, or the QSO
candidates whose spectra remains to be taken (\#19, \#36). The nearly
coincidence of the redshifts of two of these objects with the redshifts of the
HII galaxies in the filament make us think that they have likely a common
interpretation: either objects with $z$=0.245 and $z$=0.246 and objects with
$z$=0.394 and $z$=0.401 belong to the same two groups of galaxies respectively
(in a cosmological redshift interpretation) or all of them are ejected by the
parent galaxy NGC 7603 (in a non-cosmological redshift interpretation).
Nevertheless, as said in \S \ref{.clusters}, in the cosmological interpretation
it would still remain to explain the low value of $P$. Therefore, if we want to
avoid the word ``coincidence'' in all aspects (positions and redshifts)  we
must assume that all objects (\#1, \#2, \#22, \#23) are ejected by the parent
galaxy. 

HST images might show the interaction of the filament with objects \#1 and \#2 
(see \S \ref{.further1}). The narrow line character in these objects (in
principle, classified as HII galaxies according to their line ratios) would be
a result of the ejection and interaction with the filament. Evidence is shown
in other papers (e.g., Keel et al. 1998, 1999; Arp 1999a;  Burbidge
et al. 2003) that when QSOs interact with ambient material they become less
compact and had narrower lines emitted from more a more diffuse body. This
could be the physical explanation. Dynamically disturbed starburst galaxies, as
illustrated by the case of NGC 2777 (Arp 1988), tend to be the small companions
of larger  nearby galaxies belonging to older stellar populations. According to
Arp (1988), they are  recently created galaxies in which star formation is
stimulated by recent ejection from the parent galaxy; some older stars,
together with  stellar material, are suggested to be removed from the larger
galaxy in the course of this ejection.  In the system NEQ3 near NGC4151, a QSO
and an HII-galaxy have almost identical  redshifts, with a separation of 2.8
arcseconds and  nearly the same magnitudes; the HII-galaxy is embedded in a
filament while the QSO is a little bit further away (Guti\'errez \&
L\'opez-Corredoira, in preparation). It is another example of environment where
QSOs and narrow emission line galaxies have some relation.  Therefore, the
established fact of observing narrow emission line galaxies instead of QSOs is
also contemplated in the theory, although the analysis of QSO-galaxy
associations is more frequent. The origin of these sources, through the
interaction, would also explain the high observed equivalent width in their
H$\alpha $ lines.

According to this theory, the intrinsic redshifts are indicated to evolve in
discrete steps as the QSOs evolve into galaxies (Arp 1999b). The peaks in the
quantization of the redshifts would be at redshifts around 0.06, 0.30, 0.60,
0.96, 1.41, 1.96 (Arp et al. 1990; Burbidge \& Napier 2001), although the
dispersion is large  mainly because of peculiar velocities. 

The redshifts of the HII-galaxies suggest a possible relation in pairs of
objects: the pair in the filament with redshifts 0.245 and 0.394 could stem from
the same original source with intrinsic redshift $\approx 0.32$ (the exact value
of this number indeed depends  on the respective masses of the HII-galaxies), 
and a superposed Doppler radial velocities of around $\pm 17000$ km/s, for
instance (velocities of this order are obtained by Bell 2002a). A similar  pair
might be the HII-galaxies at 0.246 and 0.401 away from the filament, on the
other side of NGC 7603, but these might be well in the background.  This value
of $z\approx 0.32$ (around $z\approx 0.28$ for an observer at NGC 7603) is close
to the peak in the periodicity of QSOs/galaxies of $z=0.30$ (Arp et al. 1990;
Burbidge \& Napier 2001). The same argument might be applicable to the pair of
objects \#22-\#23. The emission in pairs or triplets could be very common
according to this theory. Bell (2002a,b) proposes that the ejection occurs in
triplets along the rotation axis of the central torus, and that these triplets are
composed of a singlet and a pair that simultaneously separate in opposite
directions and at 90$^\circ $ to the triplet ejection direction. The separation
between the singlet and the pair is higher than the pair separation; if this
were the case in NGC 7603, we would have to find the singlets somewhere in the
field of NGC 7603.

There is no unique representation of the system in terms of this theory of
ejection. We do not have enough information about the distances of the sources
with respect to the parent galaxy to build an unique 3-D representation. For
instance, Fig.~8 represents a possible configuration according to
the
ejection theory. The inclination of the galaxy is around 20 degrees with 
respect
the line of sight (ellipticity $\approx 0.35 $), so slight deviations of the
objects from the rotation axis could produce the projected image that we have
observed. Figure~8 shows a model in which the filament is not
in the plane of the galaxy, but is ejected in a direction nearly perpendicular to
the plane.

The filament does not have a blue colour like the other spiral arms in NGC 7603; 
neither does
it have young star formation since it has no emission lines (paper I); 
instead, it has
a red colour (see Fig.~3), 
like the old population of the disc of NGC 7603. Therefore, the filament 
could possibly be some material stripped from the main galaxy as a result of
some tidal interactions
or ejection. A reason for the visibility of the filament in this ejection 
with 24.0 mag/arcsec$^2$, while is not observed
in other systems, might be the integration along the line of sight of
a filament that is nearly tangential to the line of sight and provides a 
high column density.
Nonetheless, as has been said, there are some other cases which also have similar
continuous or nearly-continuous filaments/arms (with some gaps) connecting 
different-redshift objects (see \S \ref{.probab}). 
Perhaps NGC 7603 is the clearest case, but it may not be unique.

The other side of NGC 7603 (behind NGC 7603 if we assume that the filament and its
ejected objects are in front of it) could also have some ejected objects. 
We do not see the filament there, possibly because it is behind the galaxy.

Other possible scenario within this ejection hypothesis would be that all the galaxies are
in the plane of NGC 7603. It is noteworthy that all the five HII galaxies and NGC 7603B are 
almost aligned, which could lead us to think of an ejection along some common axis. However,
this axis would not be the rotation axis, which is the expected axis in ejection
 theories.

\section{Summary}
\label{.conclus}

\begin{itemize}

\item We present new observations in the field of the Seyfert galaxy NGC 7603.
These comprise broad and narrow band imaging, and intermediate
resolution spectroscopy of several objects in the field.

\item We have delineated the halo around NGC 7603 out to the isophote 26.2
mag/arcsec$^2$ in the Sloan r band  filter
finding several signs of irregularities and asymmetries towards the east.
Neither these eastern asymmetries nor the filament towards the east, apparently 
connecting NGC 7603 and NGC 7603B,
can be easily understood in an isolated galaxy, and until now no good candidates
of companions on the east side of NGC 7603 with the same redshift have been found.

\item With improved spectra with respect to those published in Paper I, we have
confirmed the redshift of the two objects in the filament connecting NGC 7603
and NGC 7603B and we have observed their $H\alpha $ lines for the first time.

\item The better resolution achieved in these new spectra and HST imaging of
these objects have allowed their more accurate classification as HII
galaxies. We have not detected any signs of variability in these objects at 
levels {\bf $\ge 0.3-0.4$} mag. We found very strong star formation (or whatever
the cause of the high equivalent widths of H$\alpha $ lines) in both
of them and the HST images show some distortions in the shape of both galaxies,
which might suggest an interaction with the filament.

\item We have detected new narrow emission line galaxies at $z=$0.246, 0.117 and
0.401, 0.8, 1.5, 1.7 minutes to the  West of the filament between NGC 7603 and NGC 7603B. 
The nearly coincidence of the redshifts of two of these objects with the
redshifts of the HII galaxies in the filament make us think that they have likely a
common interpretation: either objects with $z$=0.245 and $z$=0.246 and
objects with $z=$0.394 and $z=$0.401 belong to the same two groups of galaxies respectively
(in a cosmological redshift interpretation) or all of them are ejected by the parent
galaxy NGC 7603 (in a non-cosmological redshift interpretation).

\item The probability of a fortuitous accumulation of objects as bright as  NGC
7603, NGC 7603B, and the two objects in the filament has been computed resulting
in $\sim 3\times 10^{-9}$. The (possible, although not sure) detection of
vigorous star formation observed in the HII-galaxies of the filament, if
confirmed, would have a probability $2\times 10^{-4}$ giving a total probability
$\sim 6\times 10^{-13}$. They look dwarf HII-galaxies (non-cosmological
redshift) rather than normal/giant HII-galaxies (cosmological redshift).

\item An explanation in terms of cosmological redshifts (with or without
gravitational lensing, with or without clusters in the line of sight  has very
low probability although it is not impossible.  Alternative explanations have
been analysed.

\end{itemize}

\

{\bf Acknowledgments:}
Thanks are given to the referee Jack Sulentic for useful comments and criticisms on the interpretations
of probabilities. We thank also to 
Evencio Mediavilla (IAC) who read the manuscript and
gave helpful comments.
This research has made use of the SIMBAD database,
operated at CDS, Strasbourg, France.

\appendix

\section{Cumulative counts of galaxies and QSOs}
\label{.counts}

The cumulative counts of galaxies in the B-band can be derived from differential galaxy counts
from Metcalfe et al. (1991) for galaxies between $20.5<B<24.5$ (magnitudes 
corrected for extinction):

\[
\log N(B_0-0.25<B<B_0+0.25)
\]\begin{equation}
=0.494B_0-7.72 \ {\rm deg^{-2}}
\label{difcounts}
.\end{equation}
The cumulative count is:
\[
N(B<B_{lim})\approx 2\int _{-\infty}^{B_{lim}}dB_0N(B_0-0.25<B<B_0+0.25)
\]\begin{equation}=
3.35\times 10^{-8}\times 3.12^{B_{lim}}  \ {\rm deg^{-2}}
\label{countsgalb}
.\end{equation}
This assumes as an approximation that eq. (\ref{difcounts}) applies for $B_0<20.5$, which is more
or less correct because the change of slope is very small for lower magnitudes.

The cumulative QSO counts are given by:

\begin{equation}
N(b_j<b_{j,lim})\approx 1981-214.2b_{j,lim}+5.792b_{j,lim}^2  \ {\rm deg^{-2}}
\label{nbj}
.\end{equation}
We derived this expression by fitting the cumulative QSO counts in the
photographic $b_j$ filter from the survey by Boyle et al. (2000),  based in a
multicolour selection of QSOs.  Fig.~9 shows that fit.
The Boyle et al. (2000) data are for $b_j<21.0$, but we extrapolate the fit
(\ref{nbj}) as an approximation to higher magnitudes.  Another point obtained
from a spectroscopic survey of faint QSOs (Boyle et al. 1991) confirms that the
fit and its extrapolation are reasonably good (see Fig.~9). There are some uncertainties in these estimates but
they are low. One major concern is whether these QSO counts are complete, and we
know that they are quite complete. The multicolour selection  of QSOs is
complete for QSOs of redshift of $z<2.2$  (90\%), or 80\% for $z>2.2$ (Boyle et
al. 2000; Meyer et al. 2001). There is no selection effect that could favour the
statistics: only  10$-$15\% of extra QSOs which are not included in these
counts.

\newpage

\section*{FIGURES}

\begin{enumerate}

\item  A grey scale and contour image in the R band of the region around the 
galaxy NGC 7603. The contours correspond to isophotes 24.8, 25.3 and 26.2
mag/arcsec$^2$.

\item  A grey-scale R band image and contours corresponding
to H$\alpha$ emission at the redshift of the galaxy NGC 7603.

\item   Sloan g-r colour  of NGC7603. From bluer to redder colours [lower to higher 
values of (g-r)]:  black-blue-green-red-white.
The center of NGC~7603 is saturated. Noteworthy aspects are the 
red colour of an asymmetrical strip crossing NGC 7603, the young population (blue) at the north of NGC 7603 
and the average (green) colour of of the filament connecting NGC 7603 and NGC 7603B.

\item   Position of the sources in table \protect{\ref{ugri_table}}
(only sources with $m_u\le 23.8$ except source \#1; NGC 7603 and NGC 7603B
not included). With the double circle, we point out the three sources
which follow eq. (\protect{\ref{selecQSO}}), candidate QSOs 
by the multicolour selection. Dot-dashed lines represent the
position of the three long slits placed in the field of 
NGC~7603 to obtain the spectra of some objects.

\item   Diagrams colour-apparent magnitude and 
colour-colour of objects that were selected in
the field of NGC 7603 (table \ref{ugri_table}). The square represents object \#2.
Object \#1 is not in the plots because we have not its u magnitude.
The lines indicate the limits of (g-r) colour under which 
QSOs are likely to be found (see \S \ref{.colours}): there are 
3 candidates with (g-r) under the red line.

\item   HST image in the F606W filter  of the region centred on the filament between
NGC 7603 and NGC 7603B. Also shown are the contours of the two objects in the
filament. Note that there are many bad pixels/cosmic rays in the images that do not correspond to any
object. The PSF is $\sim 0.1$ arcsec. Dotted lines show the area (around
140 arcsec$^2$) that we consider ``filament'' for the calculation of the
probabilities in \S \protect{\ref{.probab}}.

\item   Main spectral features (corrected of redshift) of objects \#1
($z=0.245$, in the filament), \#2 ($z=0.394$, in the filament), \#21
($z=0.117$),  \#23 ($z=0.246$) and \#22 ($z=0.401$). Dashed lines below the
spectra are their zero-flux levels.

\item   Possible representation of the system of NGC 7603/NGC 7603B/Object \#1/Object 
\#2 if we accept the hypothesis of the three last objects to be ejected by 
the parent galaxy,
NGC 7603. The inclination of NGC 7603 with respect the line of sight is 
$\approx 20^\circ $. The major axis  in the projected image (Axis 1) has a 
position angle $\approx -15^\circ $; the minor axis in the projected image is ``Axis 2''.

\item   Cumulative QSO counts data (Boyle et al. 2000; 1991)
and a fit of a second polynomial degree to the Boyle et al. (2000) data.

\end{enumerate}

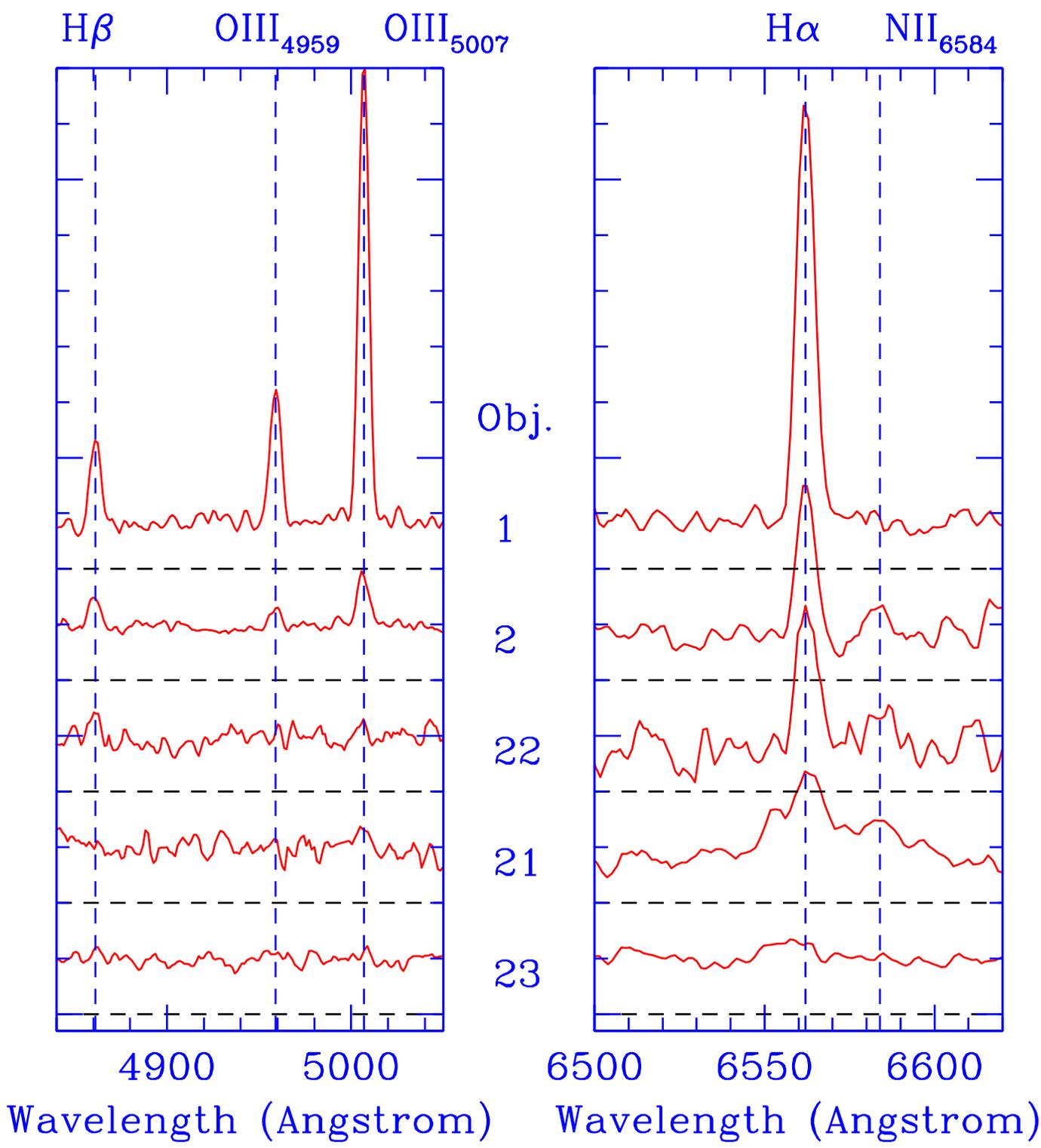
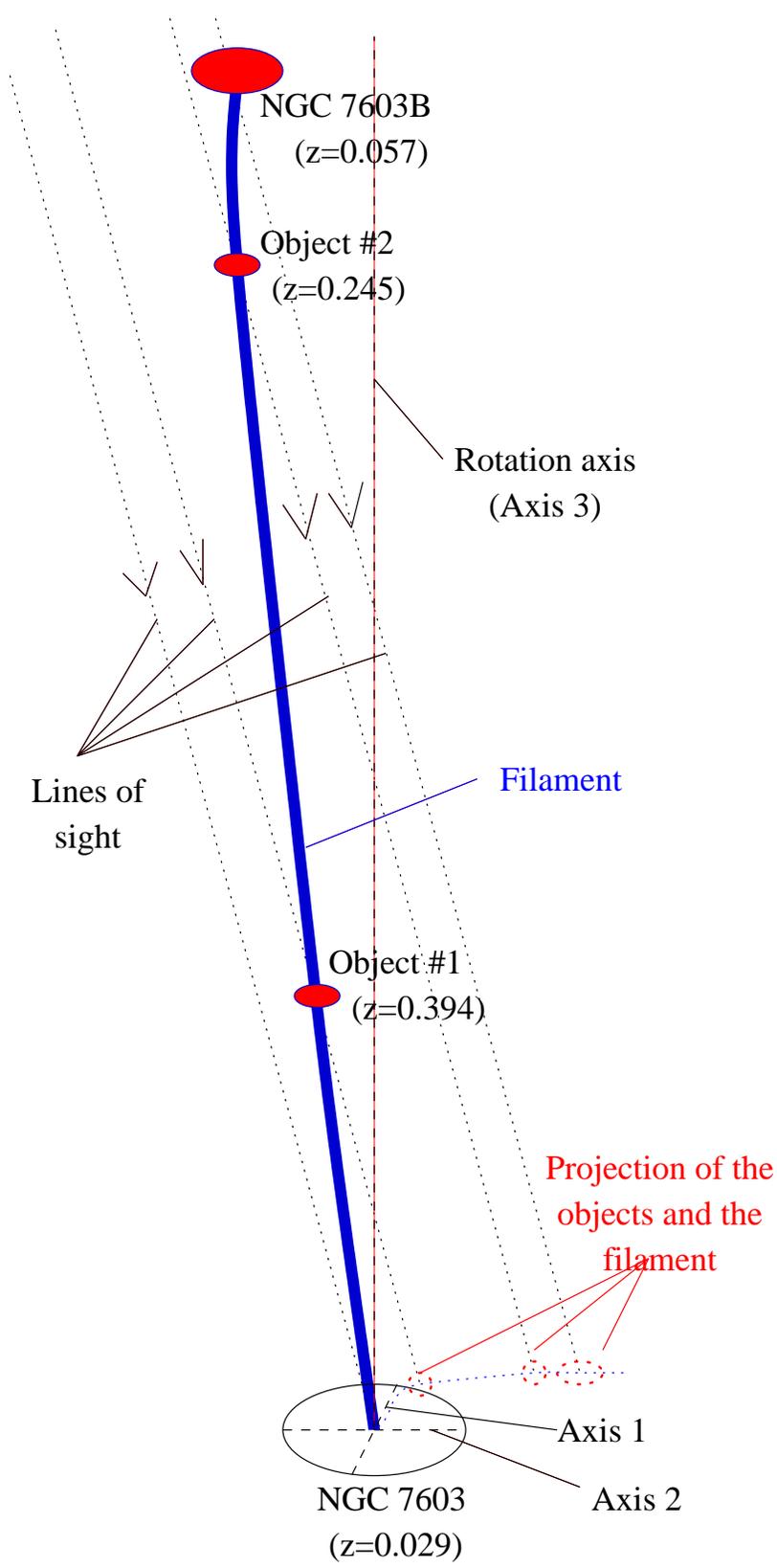
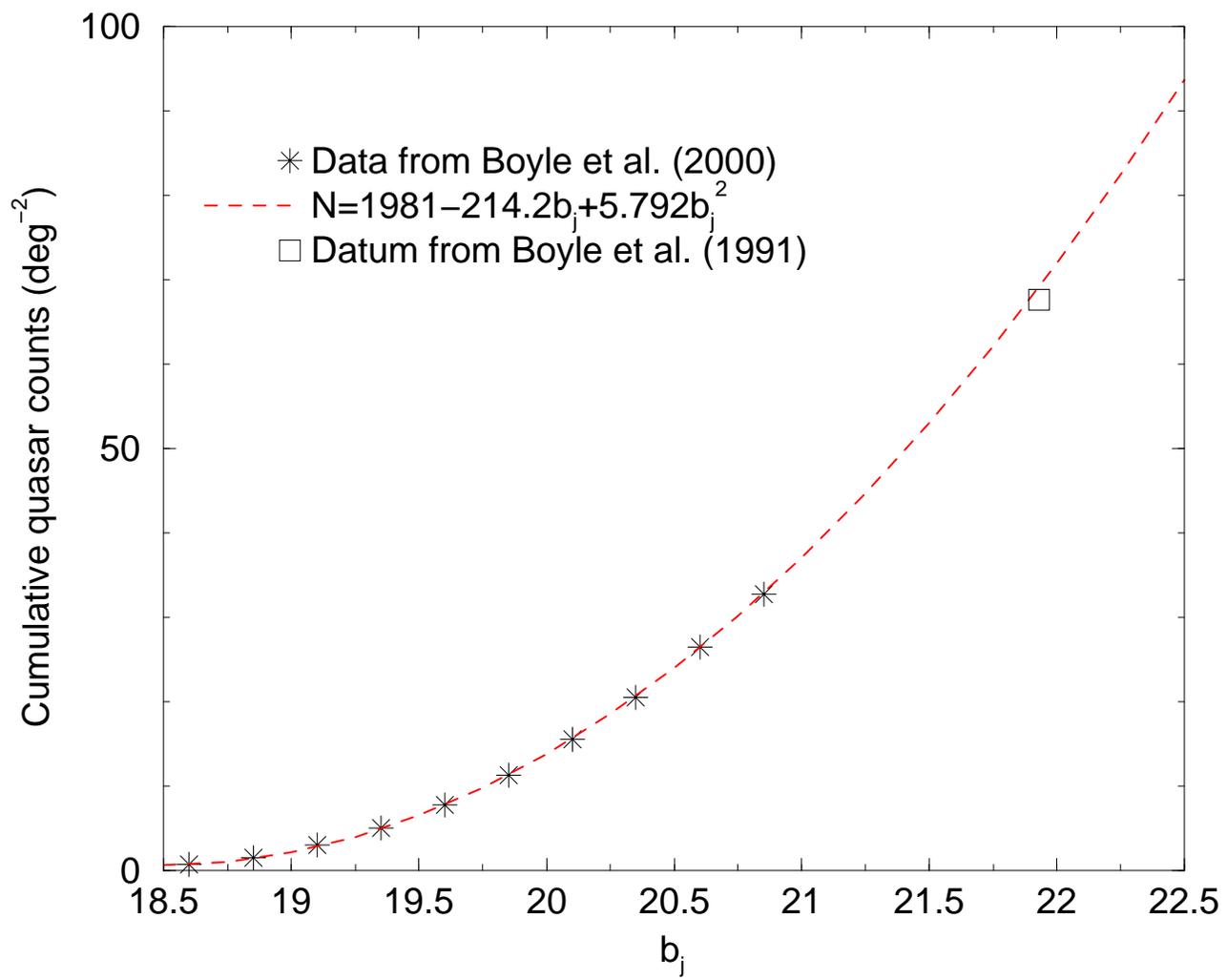
\end{document}